\documentclass[conference]{IEEEtran}
\IEEEoverridecommandlockouts

\usepackage{cite}
\usepackage{amsmath,amssymb,amsfonts}
\usepackage{algorithmic}
\usepackage{graphicx}
\usepackage{textcomp}
\usepackage{xcolor}
\usepackage[hyphens]{url}
\usepackage{caption}
\usepackage{hyperref}

\usepackage{geometry}
\usepackage[all=normal,paragraphs=tight,floats=tight,mathspacing=tight,wordspacing=tight,leading=tight,mathdisplays=tight]{savetrees}
\usepackage[stretch=0,shrink=100]{microtype}

\usepackage{setspace}
\begin{document}

\pdfpagewidth=8.5in
\pdfpageheight=11in

\pagenumbering{gobble}
\pagestyle{empty}

\title{\huge HLSFactory-Agent: Large-Scale Agentic HLS Dataset Construction from Academic and Open-Source Projects\vspace{-1ex}}

\author{
\IEEEauthorblockN{Kaushik Chandana$^{1,*}$, Jay Imperatori$^{1,*}$, Tanmay Shukla$^{1,*}$, Justin Zhou$^{1,*}$, Stefan Abi-Karam$^{1,2}$, Callie Hao$^{1}$%
\thanks{$^{*}$Equal Contribution}}
\IEEEauthorblockA{$^{1}$Georgia Institute of Technology, Atlanta, USA \quad $^{2}$Georgia Tech Research Institute, Atlanta, USA}
\IEEEauthorblockA{\{kchandana6, jimperatori6, tshukla31, justin.zhou, stefanabikaram, callie.hao\}@gatech.edu}
}

\maketitle

\thispagestyle{plain}
\pagestyle{plain}

\begin{abstract}
Building large, diverse datasets of high-level synthesis (HLS) designs beyond common community benchmarks remains an open challenge. This challenge is made urgent by the rise of deep learning and LLMs for hardware design, which demand such datasets to train QoR models and benchmark LLMs on HLS tasks. Despite ongoing efforts to broaden sources, dataset curation still depends on manual work: locating HLS designs across academic publications and open source, then extracting standalone designs from larger codebases. The process is error-prone and demands expert knowledge, iterative testing, and substantial per-repository engineering.

To address this, we present HLSFactory-Agent, an LLM agent that automates large-scale HLS dataset curation by extracting standalone designs from larger codebases. HLSFactory-Agent runs the open-source Pi agent framework inside Docker containers to build and evaluate each extracted design. This turnkey automation allows users to pass a GitHub link or code directory to HLSFactory-Agent and receive a folder of extracted HLS designs ready to be integrated into the HLSFactory dataset framework. Additionally, we provide open-source scripts to scrape and index papers from computer architecture, EDA, and FPGA conferences that possibly implement or use HLS designs, allowing for faster human discovery and curation of HLS designs for HLSFactory-Agent.

We report initial results from running HLSFactory-Agent across a small subset of our indexed repositories, demonstrating successful extraction of synthesizable designs from structured codebases.

We open source HLSFactory-Agent and indexing scripts at \texttt{\href{https://github.com/sharc-lab/hlsfactory-agent}{github.com/sharc-lab/hlsfactory-agent}}.

\end{abstract}

\section{Introduction}

High-level synthesis (HLS) enables domain-specific hardware experts to build flexible, high-performance hardware accelerators. Recent advances in deep learning and large language models have accelerated the HLS design workflow through quality-of-results prediction models, fast design space exploration tools, and LLM-based frameworks for iterative design and optimization. In parallel, compiler engineers have continuously improved HLS compilers to provide higher-level design abstractions and faster compilation.

These innovations depend critically on access to large and diverse HLS design datasets and benchmarks. While popular community benchmarks exist, they remain limited in diversity and size, prompting efforts to expand HLS design collections and build community dataset infrastructure. HLSFactory \cite{hlsfactory}, our prior open-source framework, addresses this need by providing infrastructure to curate HLS designs and automatically generate synthesis and implementation data as complete packaged datasets for downstream research. Similarly, evaluating large language models and agentic coding tools for HLS requires robust benchmark collections. Our prior work, HLS-Eval \cite{hlseval}, provides this infrastructure with baseline evaluations for HLS code generation and editing tasks. However, both HLSFactory and HLS-Eval face a common bottleneck: manually curating HLS designs from academic publications and open-source repositories is labor-intensive, error-prone, and requires substantial per-repository engineering effort.

To address this challenge, we present HLSFactory-Agent, an AI agent that automates large-scale HLS dataset curation by extracting standalone, synthesizable designs from unstructured academic and open-source repositories. Additionally, we provide open-source scripts to scrape and index HLS-related papers from major computer architecture, EDA, and FPGA conferences, enabling rapid discovery of candidate repositories for automated extraction.

\begin{figure*}[t!]
    \centering
    \includegraphics[width=\linewidth]{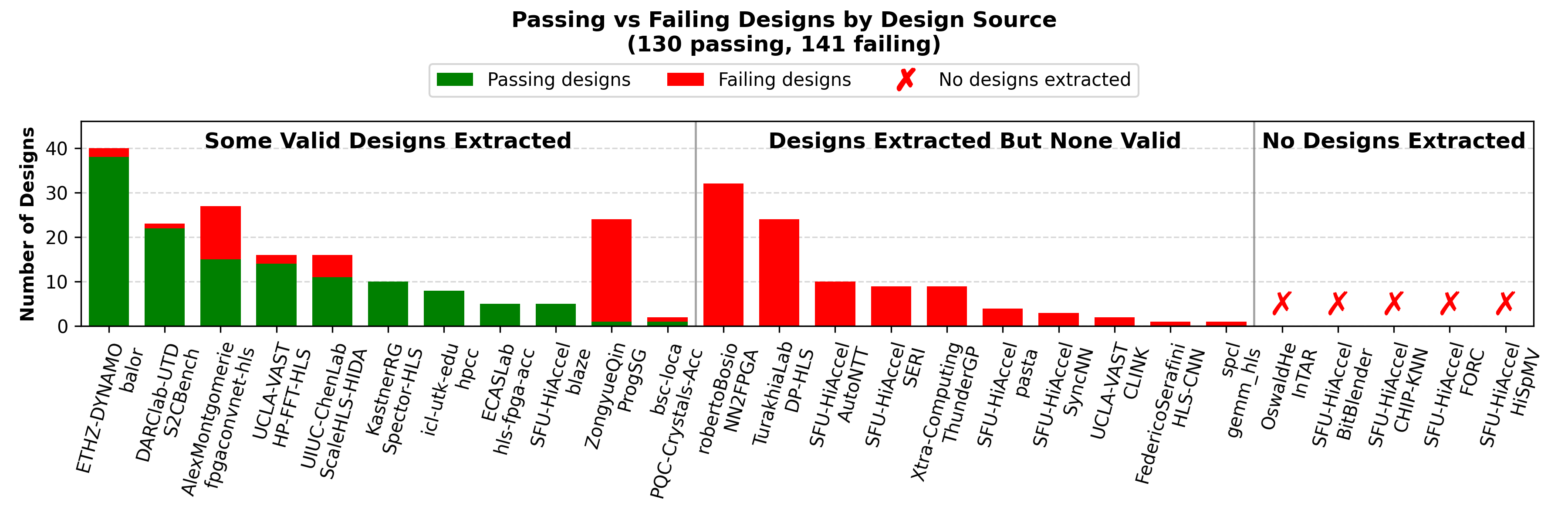}
    \vspace{-2em}
    \caption{Passing and failing extracted designs by source repository. Green bars count candidate designs whose generated \texttt{synth.tcl} completes Vitis HLS synthesis, red bars count extracted candidates that fail validation, and red crosses mark repositories where the agent produced no candidate design folders. Our evaluation run extracts $271$ candidate designs across $26$ repositories: $130$ passing and $141$ failing.}
    \label{fig:recovered_designs}
\end{figure*}

\section{Methodology}

\subsection{Design Curation from Academic Publications and Open-Source Code Repositories}

The computer architecture, FPGA, and EDA communities have open-sourced many HLS designs on platforms like GitHub. We propose scraping relevant publications and finding their associated code repositories.

Our source-indexing workflow searches for "high-level synthesis," "high level synthesis," and "HLS" keywords across:
\begin{itemize}
    \item FPGA venues: FPGA, FCCM, FPL, FPT, HEART, TRETS
    \item EDA venues: DAC, ICCAD, ASP-DAC, DATE, MLCAD, GLVLSI, HOST, TCAD
    \item Architecture venues: ISCA, MICRO, HPCA, ASPLOS, ESWEEK, MLSys
\end{itemize}

We provide scripts that search publications, merge, and deduplicate entries using keyword matching of titles and abstracts. The repository includes scrapers for DBLP, IEEE Xplore, and Crossref/ACM metadata. Deduplication uses DOI extraction when available and normalized title/year matching otherwise. Output is a CSV/Excel file of candidate papers with metadata and links.

We collect 2,517 candidate publications (from between 1984--2026) potentially implementing HLS designs. These publications still require human review; some papers use HLS peripherally, some artifacts are unavailable, and some repositories contain generated or vendor-specific code difficult to redistribute. Human curation identifies papers with public repositories, enabling efficient filtering for HLSFactory-Agent design extraction.

\subsection{Agentic Design Extraction from Codebases}

HLSFactory-Agent is our LLM agent that automatically extracts HLS designs from unstructured repositories. Built on the open-source Pi agent framework \cite{pi}, chosen for its simplicity and structured session logs that enable detailed analysis, it runs inside an Ubuntu Docker container via a Python API for isolation and reproducibility. This setup gives the LLM agent access to a C++ compiler (Clang), file reading, writing, and editing tools, and Bash. We find this minimal toolset sufficient for the agent to self-test extracted designs. We assume that designs compilable with Clang have a high chance of synthesizing with Vitis HLS, since HLS synthesis flows begin with C++ compilation.

The agent prompt outlines the following extraction pipeline:

\begin{enumerate}
    \item Analyze the repository and identify all HLS designs or top-level functions
    \item Create one output directory per design under \texttt{output\_hls\_designs/<design\_name>/}
    \item Copy all required source files, headers, utilities, and test data into each design folder
    \item Convert non-Vitis constructs into Vitis-compatible forms (\texttt{ap\_int}, \texttt{ap\_fixed}, \texttt{hls::stream})
    \item Preserve existing testbenches or generate new ones when none exist
    \item Compile source files with Clang in syntax-only mode using the Vitis include path
    \item Generate a \texttt{synth.tcl} file with top-level function, source list, testbench list, target part, and \texttt{csynth\_design}
\end{enumerate}

HLSFactory-Agent records the full agent session in a JSONL trace and exports an HTML transcript for debugging and analyzing agent behavior, tool usage, runtime, and token cost.

\section{Initial Results}

We evaluate HLSFactory-Agent using the \texttt{DeepSeek-V4-Flash} model on an initial set of 26 public repositories drawn from academic and open-source HLS projects. The set includes projects such as \texttt{fpgaconvnet-hls}, \texttt{S2CBench}, \texttt{balor}, \texttt{DP-HLS}, \texttt{HP-FFT-HLS}, \texttt{CLINK}, \texttt{ThunderGP}, \texttt{gemm\_hls}, \texttt{NN2FPGA}, and multiple \texttt{SFU-HiAccel} repositories. We process each repository independently with its own HLSFactory-Agent launched in parallel. We then validate each extracted design using Vitis HLS to synthesize each extracted candidate design. We report a design as "valid" or "passing" only if the generated \texttt{synth.tcl} runs HLS synthesis and completes successfully. This is a strict check: designs that contain useful extracted code but fail due to Tcl errors, missing headers, target part issues, unsupported constructs, or timeouts count as failing.

Across the 26 repositories, HLSFactory-Agent extracts 271 candidate designs. Of these, 130 pass Vitis HLS synthesis, and 141 fail. Five repositories produce no extracted designs in this initial run. Figure \ref{fig:recovered_designs} summarizes passing and failing designs by source repository. Several repositories yield many passing designs, while others yield candidate designs but no passing designs, indicating that the agent found plausible kernels but did not extract them into valid standalone Vitis HLS designs. A final group of repositories yields no designs, either because the agent could not identify suitable kernels or because the repository structure was incompatible with the current prompt.

\begin{figure}[th]
    \centering
    \scalebox{1}[1.0]{\includegraphics[width=\linewidth]{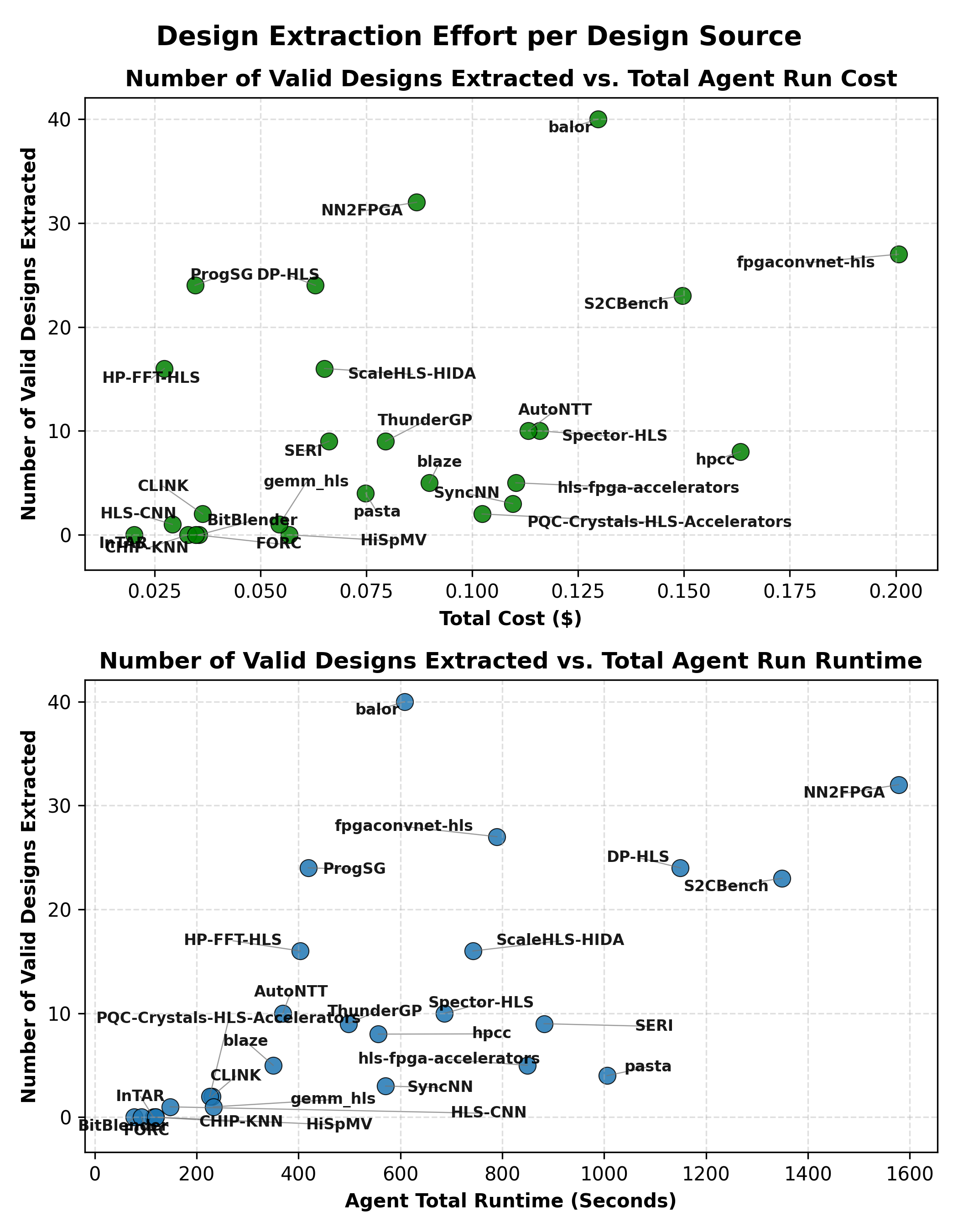}}
    \vspace{-1.6em}
    \caption{Extraction effort per design source. Each point is one repository run. The positive relationship suggests inference-time scaling: more designs extracted require longer, costlier runs.}
    \label{fig:effort}
\end{figure}

Agent traces also reveal the extraction effort per source. Figure \ref{fig:effort} plots extracted designs against model inference cost and agent runtime. Repositories yielding more designs require more inference effort. Some projects produce many designs cheaply because of regular structure and easily identified kernels. Others require more runtime for comparable yields because of complex dependencies, generated code, or build assumptions. This suggests that we should evaluate agentic extraction by designs per dollar spent and designs per minute run, not just by total extracted dataset size.

\section{Ongoing / Future Work}
As part of ongoing work, we aim to further automate the paper-by-paper review and associated code repository discovery using LLMs to review papers for open code repositories or search for associated repos, rather than relying on human review. Furthermore, we are working to allow HLSFactory-Agent to run Vitis HLS Synthesis within the Docker container to add another layer of verification feedback to the agent and catch more failing cases. Finally, as we extend HLSFactory-Agent and gather new designs, we plan to contribute these designs back to HLSFactory and HLS-Eval benchmarks, enriching design resources for the HLS community.

\bibliographystyle{IEEEtranS}
\bibliography{refs}

@inproceedings{hlsfactory,
	location = {Salt Lake City {UT} {USA}},
	title = {{HLSFactory}: A Framework Empowering High-Level Synthesis Datasets for Machine Learning and Beyond},
	isbn = {979-8-4007-0699-8},
	url = {https://dl.acm.org/doi/10.1145/3670474.3685961},
	doi = {10.1145/3670474.3685961},
	shorttitle = {{HLSFactory}},
	eventtitle = {{MLCAD} '24: 2024 {ACM}/{IEEE} International Symposium on Machine Learning for {CAD}},
	pages = {1--9},
	booktitle = {Proceedings of the 2024 {ACM}/{IEEE} International Symposium on Machine Learning for {CAD}},
	publisher = {{ACM}},
	author = {Abi-Karam, Stefan and Sarkar, Rishov and Seigler, Allison and Lowe, Sean and Wei, Zhigang and Chen, Hanqiu and Rao, Nanditha and John, Lizy and Arora, Aman and Hao, Cong},
	urldate = {2026-03-24},
	date = {2024-09-09},
	langid = {english},
}

@inproceedings{hlseval,
	location = {Stanford, {CA}, {USA}},
	title = {{HLS}-Eval: A Benchmark and Framework for Evaluating {LLMs} on High-Level Synthesis Design Tasks},
	rights = {https://doi.org/10.15223/policy-029},
	isbn = {979-8-3315-2597-2},
	url = {https://ieeexplore.ieee.org/document/11106033/},
	doi = {10.1109/ICLAD65226.2025.00021},
	shorttitle = {{HLS}-Eval},
	eventtitle = {2025 {IEEE} International Conference on {LLM}-Aided Design ({ICLAD})},
	pages = {219--226},
	booktitle = {2025 {IEEE} International Conference on {LLM}-Aided Design ({ICLAD})},
	publisher = {{IEEE}},
	author = {Abi-Karam, Stefan and Hao, Cong},
	urldate = {2026-03-24},
	date = {2025-06-26},
}

@software{pi,
	title = {earendil-works/pi},
	rights = {{MIT}},
	url = {https://github.com/earendil-works/pi},
	publisher = {Earendil Works},
	urldate = {2026-05-13},
	date = {2026-05-13},
	note = {original-date: 2025-08-09T14:03:50Z},
}

\end{document}